# The Formation and Evolution of Planetary Systems: The Search for and Characterization of Young Planets

**Charles Beichman** (NExScI), **Isabelle Baraffe** (ENS, Lyon), **Christopher Crockett** (UCLA/Lowell)**, Sarah Dodson-Robinson** (NExScI), **Jonathan Fortney** (UCSC), **Andrea Ghez** (UCLA), **Thomas P. Greene** (NASA Ames), **Adam Kraus** (Caltech), **Doug Lin** (UCSC), **Naved Mahmud** (Rice), **Fabien Malbet** (Univ. Grenoble), **Mark Marley** (NASA Ames), **Rafael Millan-Gabet** (NExScI), **Lisa Prato** (Lowell Observatory), **Ben Oppenheimer** (AMNH), **Michal Simon** (SUNY Stony Brook), **John Stauffer** (IPAC), **Angelle Tanner** (JPL) **, T. Velusamy (JPL)**

*Contact Information for Primary Author:*

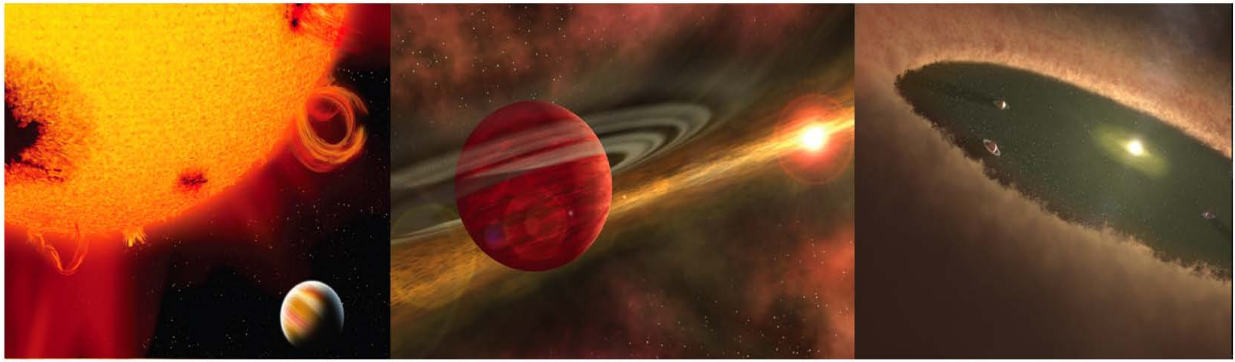

*C. Beichman, NASA ExoPlanet Science institute*
*818-653-8220, chas@ipac.caltech.edu*
*February 17, 2009*

Abstract

*Despite the revolution in our knowledge resulting from the detection of planets around mature stars, we know almost nothing about planets orbiting young stars because rapid rotation and active photospheres preclude detection by radial velocities or transits and because direct imaging has barely penetrated the requisite range of high contrast and angular resolution. Thus, our knowledge about the formation of planetary systems is rudimentary at best. Among the key questions that can be addressed in the coming decade are competing theories of planet formation, e.g. core accretion vs. disk fragmentation; the relationship between emergent intensity and total luminosity as a function of planetary mass and age; the effects of tidal and orbital migration on solar system architecture; and the formation of planets as a function of disk and stellar mass.*

*While radial velocity and transit observations may be able to detect a few planets, astrophysical effects will limit these techniques to relatively high mass, small orbital radius systems. Of the techniques presently under consideration for the coming decade, only space-based astrometry offers the prospect of discovering gas giants (100 to >> 300 $M_\oplus$), lower mass systems such as icy giants (10 to 100 $M_\oplus$), and even a few rocky, super-Earths (~10 $M_\oplus$) orbiting stars ranging in age from 1 to 100 Myr. Astrometry will complement high contrast imaging from large ground-based telescopes and JWST which should be able to detect gas giants (1~10 $M_{Jup}$) in orbits from a few to a few hundred AU. An astrometric survey in combination with imaging data for a subsample of objects will allow a detailed physical understanding of the formation and evolution of young gas giant planets impossible to achieve by any one technique. In this white paper we focus on astrometric surveys for young planets.*

## I. Important Scientific Questions in Planet Formation

Exoplanets are numerous and diverse with "hot" and "cold" Jupiters, Neptunes, and Super-Earths now being detected around mature stars with great regularity. But we do not know how planets form with such a wide diversity of total masses, core radii, total radii and orbital properties. During the next decade we will able to establish the initial conditions of the gas and dust in protoplanetary disks as well as to detect young planets to investigate their formation, migration, and early evolution. This white paper focuses on *the incidence and properties of young planets* and complements the white paper prepared by Millan-Gabet et al. which discuses high angular and spectral resolution imaging of disks.

Many critical questions concerning the formation and early evolution of gas and ice giant planets suffer from a near-total lack of data about the properties of planets orbiting young stars. What processes affect the formation and dynamical evolution of planets? When and where do planets form? What is the initial mass distribution of planetary systems around young stars? How might planets be destroyed? What is the origin of the eccentricity of planetary orbits? What is the origin of the apparent dearth of brown dwarf companions to mature stars? What accretion mechanisms might explain the apparent gap between the critical core mass and planets of Jupiter-Saturn mass? How might the formation and migration of gas-giant planets affect the formation of terrestrial planets? How do the observable properties of a planet change with its mass and evolve with time? Two overarching questions concerning the formation process itself and the subsequent evolution of individual gas and icy giant planets can be addressed directly in the coming decade.

### A. How do Planets Form: Core accretion vs. Disk Fragmentation?

The core accretion model of planet formation has had a number of successes in explaining the dependence of both the incidence of planets and the masses of their cores as a function of stellar metallicity (Ida and Lin 2005; Fischer and Wright 2005; Liu, Burrows and Ibgui 2008). These results are naturally explained via a mechanism whereby a solid core built of refractory elements gathers gaseous material from the protostellar disk. This mechanism has difficulties explaining the existence of planets located far from the host star where timescales are long and the disk surface density low. Despite improvements to the model based on an enhanced surface density due to enrichment by ices (Robinson 2008), an alternative scenario of multiple fragmentation events in a massive disk may be needed to form more distant planets (Boss 2001), including those imaged recently with HST and ground-based adaptive optics. The massive planets on wide (20-100 AU) orbits circling the young A stars Fomalhaut (Kalas et al. 2009) and HR 8799 (Marois et al. 2009) present strong challenges to planet formation theory since both the core accretion and gravitational instability mechanisms are only marginally capable of forming these planets, even under artificially favorable conditions (Kratter, Dodson-Robinson et al. , in prep). Planet searches across a range of stellar mass and orbital radii will be necessary to resolve the theoretical debate about planet formation mechanisms, e.g. whether the maximum planet mass increases with both stellar mass and orbital radius.[1]

Theoretical progress will require a survey of young stars looking for planets across a broad range of stellar mass (0.2 to 2 $M_o$), planet mass (0.1 to 10 $M_{Jup}$) and orbital properties such as semi-major axis and eccentricity. Ideally, a complete survey would cover a range from 0.1 -1 AU to address tidal migration, orbital circularization, and planet destruction; from 1-10 AU to address planet formation in and around the snowline (Pollack et al. 1996); and from 10 to beyond 100 AU to address whether distant planets formed in situ or migrated outward due to multi-planet interactions or disk dynamical effects. Our only chance for finding planets at large orbital radii will be in young systems where self-luminous planets are bright enough to be detected directly; stellar reflex motions will be too small or too slow for effective detection. As will be discussed below a combination of astrometry and imaging will be required to execute the census needed to advance our understanding of the origins of planetary systems.

### B. How Do Planets Evolve as a Function of Mass and Age?

The poor state of evolutionary tracks for young, low mass stars is well known (Baraffe et al. 2002) and extends into the giant planet regime. While numerous models predict the luminosity, effective temperature, and emergent spectrum of planets as a function of mass, age and stellar insolation (Burrows 1997; Chabrier et al. 2003; Fortney et al. 2008), these models have confronted real data only in the case of mature, transiting hot Jupiters. Almost nothing is known about purely self-luminous objects due to a lack of either imaging or dynamical observations. The only self-luminous objects presently susceptible to study are young objects found by direct imaging on distant orbits for which no dynamical mass estimate is at present possible. Indeed, mass estimates for these systems are highly uncertain and, since they rely on untested and controversial models, cannot be used to improve the models themselves. As an example of the range of this theoretical uncertainty, consider the assertion by Marley et al. (2007) that the brightness of young planets of a given age and mass may be overestimated by factors of 10 to 100 due to improper treatment of the accretion shock in the early stages of planet formation.

---

[1] For planets growing by core accretion in passively irradiated disks, the planet mass scales as $M_p \propto M_*^{25/16} a^{3/4} R_*^{-1/2}$, where $M_*$ is the star mass, $a$ is the semi-major axis (out to an effective outer disk radius of about 25 AU) and $R_*$ is the star radius (Dodson-Robinson et al. , in prep).

While observations of transiting planets around mature stars have revealed a great deal about the *end-state* of the planet formation process, many questions remain about the mass-density relationship, the size of the refractory core, and influence of insolation on planet radius (Charbonneau et al. 2007; Seager et al. 2007). To complete this picture we need information on the *initial-state* of the planet formation process. Advances in high spatial and spectral resolution observations of planet-forming disks are addressed in the Millan-Gabet White Paper. We focus here on finding and measuring the properties of individual planets with ages less than ~100 Myr.

From the combination of astrometry and imaging we can obtain masses and orbits as well as effective temperatures, luminosities (and thus derived radii) for young planets. These data will help anchor new models for the internal structure of giant planets and yield a new generation of evolutionary models, e.g. Mass-Luminosity-Teff tracks, useful for determining ages and/or masses for objects for which only imaging data are available. Spectroscopy combined with astrometric masses would also allow one to look for trends in composition with mass and orbit. The giant planets in our solar system are enhanced in heavy elements over solar abundance with Jupiter at ~4x, Saturn at ~10x, and Uranus and Neptune at factors of ~30 to 50x. This is a fingerprint of the formation process, the interpretation of which people argue about in the solar system. Similar measurements for other systems would go a long way towards understanding giant planet formation.

## II. The Challenge and Promise of Detecting Young Planets

### A. Astrometry

Astrometry promises to both find young planets undetectable by other means as well as to provide dynamical masses for objects detected by direct imaging. Astrometric surveys of young stars will probe the critical region between 1-5 AU where gas giants are thought to form, but where imaging techniques are hardest pressed to detect objects of <5 $M_{Jup}$. A census of young planets will address basic questions of planet formation and migration. The dynamical masses combined with imaging information will put evolutionary models on a sound physical footing.

|  | 1 *M* Star | | | | 0.15 *M* Star | | | |
|---|---|---|---|---|---|---|---|---|
|  | Distance, pc | | | | Distance, pc | | | |
|  | 30 pc | | 140 pc | | 30 pc | | 140 pc | |
|  | Orbit, AU | | Orbit, AU | | Orbit, AU | | Orbit, AU | |
| *Planet* | 1 | 5.2 | 1 | 5.2 | 1 | 5.2 | 1 | 5.2 |
| Jupiter, 1 $M_{Jup}$ | 32 | **170** | 7 | 36 | **214** | **1110** | 46 | **240** |
| Saturn, 0.28 $M_{Jup}$ | 9 | 47 | 2 | 10 | **60** | **311** | 13 | **67** |
| Uranus, 0.023 $M_{Jup}$ | 0.7 | 4 | 0.2 | 0.8 | 5 | 26 | 1.1 | 6 |

Table 1. Astrometric signal (μas) from planets at various distances, orbital locations, and stellar host mass. Entries indicated in bold (≥ 50 μas) would be detectable with Gaia or ground-based interferometers. Other systems (shown in green) are detectable with a mission having SIM Lite's capabilities.

As a consequence of the limits and selection biases of the radial velocity, transit, and direct imaging techniques, we know almost nothing about the incidence of planets around young stars, leaving us with many questions about the formation and evolution of gas giant and icy planets. Precision astrometry can remedy this situation. Table 1 gives typical astrometric signals for gas giants (Jupiter and Saturn) and icy giants (Uranus) at two orbital distances (1 and 5.2 AU), two distances from the Sun (140 and 30 pc, representative of 2- to 10-Myr-old stars and 10- to 100-Myr-old objects, respectively), and orbiting 0.15 and 1.0 $M_\odot$ stars. The signatures cover a range of <1 to 1,000 µas. A Jupiter orbiting 1 AU away from a 1 $M_\odot$ star at the distance of the youngest stellar associations (1 to 10 Myr) such as Taurus (140 pc) would produce an astrometric amplitude of 7 µas and up to 25 µas at the 25 to 50 pc distance of the nearest young stars (10 to 100 Myr). Using a simple model for the effect of starspots on the stellar photocenter (Beichman 2001; Tanner et al. 2007), the astrometric jitter for a typical T Tauri star at 140 pc is less than 3 µas (1 σ) for R-band variability less than 0.05 mag. Thus, the search for Jovian planets is plausible for young stars less variable than ~0.05 mag in the visible. If stellar jitter follows the $t^{-0.5}$ dependence seen for calcium plage activity (Skumanich 1972) then a 40-Myr-old, 0.15 $M_\odot$ star at a distance of 30 pc would have an astrometric jitter at <0.5 µas making planets of super-Earth to Uranus masses detectable over the semi-major axis range of 1 to 5 AU.

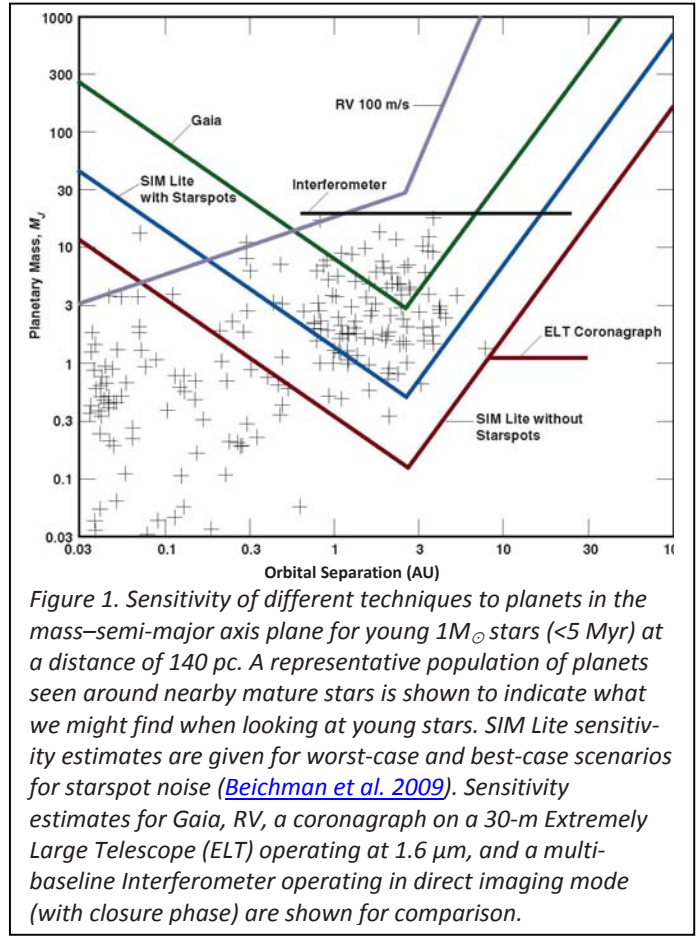

Figure 1. Sensitivity of different techniques to planets in the mass–semi-major axis plane for young $1M_\odot$ stars (<5 Myr) at a distance of 140 pc. A representative population of planets seen around nearby mature stars is shown to indicate what we might find when looking at young stars. SIM Lite sensitivity estimates are given for worst-case and best-case scenarios for starspot noise (Beichman et al. 2009). Sensitivity estimates for Gaia, RV, a coronagraph on a 30-m Extremely Large Telescope (ELT) operating at 1.6 µm, and a multi-baseline Interferometer operating in direct imaging mode (with closure phase) are shown for comparison.

ESA's Gaia mission and ground-based telescopes or interferometers (ESO's VLTI and the Keck Astra interferometer) have astrometric limits around 50 µas, almost two orders of magnitude worse than the levels possible with a targeted space interferometer. While these instruments will detect planets in nearby systems, their mass limits will be high (1 to a few $M_{Jup}$) and their target stars limited to ages greater than disk lifetimes of ~10 Myr.

### B.    Complementary Astrometry And Imaging Searches

A few objects of potentially planetary mass have been detected at 20 - 100 AU from young (<10 Myr) host stars by coronagraphic imaging, e.g., 2MASSW J1207334 (Chauvin et al. 2005), GQ Lup (Neuhauser et al. 2005) and most recently Fomalhaut and HR8799 (Marios et al. 2009; Kalas et al. 2009). However, the masses of these objects are poorly known and it is uncertain whether

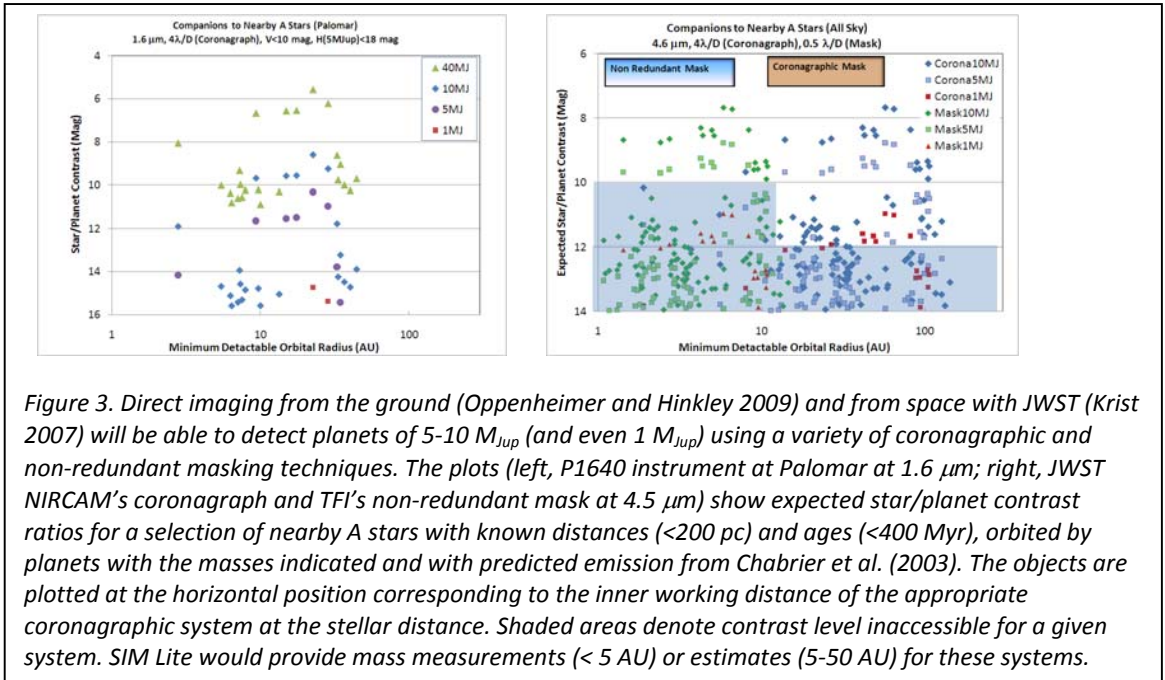

*Figure 3. Direct imaging from the ground (Oppenheimer and Hinkley 2009) and from space with JWST (Krist 2007) will be able to detect planets of 5-10 $M_{Jup}$ (and even 1 $M_{Jup}$) using a variety of coronagraphic and non-redundant masking techniques. The plots (left, P1640 instrument at Palomar at 1.6 μm; right, JWST NIRCAM's coronagraph and TFI's non-redundant mask at 4.5 μm) show expected star/planet contrast ratios for a selection of nearby A stars with known distances (<200 pc) and ages (<400 Myr), orbited by planets with the masses indicated and with predicted emission from Chabrier et al. (2003). The objects are plotted at the horizontal position corresponding to the inner working distance of the appropriate coronagraphic system at the stellar distance. Shaded areas denote contrast level inaccessible for a given system. SIM Lite would provide mass measurements (< 5 AU) or estimates (5-50 AU) for these systems.*

these are truly planets and not brown dwarfs.

As other white papers will describe, we can expect significant progress in direct imaging with high contrast, high angular resolution systems on existing and future telescopes at 1- 3 μm (Gemini, VLT, Palomar, TMT/GMT/EELT), with multi-baseline interferometers (CHARA, VLTI), and with JWST at 3-5 μm (especially with its non-redundant mask coronagraph operating with an inner working angle of a few tenths of an arcsecond; Sivaramakrishnan et al. 2009). As Figure 3 suggests, direct imaging instruments planned for the next decade can credibly claim to be able to detect 5-10 $M_{Jup}$ objects (including, perhaps, a few 1 $M_{Jup}$ objects) within 5 AU of nearby, young AFGK stars (50-1,000 Myr; 15-50 pc) and intrinsically brighter 1-5 $M_{Jup}$ objects within 5-10 AU of the youngest T Tauri stars. But direct imaging observations will not reach down to the lower masses or inner orbital radii possible with a space astrometric mission. Not will direct imaging provide dynamical masses. However, the combination of astrometry and imaging, where possible, and the imaging of planets on radii beyond the reach of astrometry will provide powerful complements to astrometric surveys.

It is worth noting that while *precise* mass measurements will not be possible for the furthest-out planets, measurements of orbital accelerations over a 5-10 year baseline can give a valuable indicator of planet mass. The deviation of a long period orbit from stellar proper motion is given by ~ *4.7 (10 pc/dist) ($M_p/M_{Jup}$) (100 AU/a)$^2$ (t/ 5yr)$^2$* μas, where *a* is the semi major axis and *t* is the measurement duration, or roughly 5 μas for a 5 $M_{Jup}$ planet in a 100 AU circular orbit around a star at 50 pc. This deviation from proper motion would be detectable by SIM-Lite and would constrain the masses of planets (or brown dwarfs!) detected with imaging.

### C. Radial Velocity

Many young stars are characterized by weak spectral features due to veiling, rotationally broadened line widths >> 1 km s$^{-1}$, large-scale radial velocity (RV) motions, and/or brightness fluctuations of many percent from starspots (Carpenter et al. 2001). Thus, visible radial velocity measurements and photometric observations have been unable to detect planets around young stars. Setiawan et al. (2008) recently claimed an RV detection of a planet orbiting the young star TW Hya, but this claim has been called into serious question (Huélamo et al. 2008) as being due to large-scale photospheric variations. Similarly, Prato et al. (2008) initially identified potential "hot Jupiters" orbiting DN Tau and V836 Tau based on visible spectroscopy, but used follow-up IR spectroscopy to demonstrate that the RV variations were due to photospheric variability, not planets. The near-IR is, however, a more promising wavelength region for RV and photometric searches of young stars due to the 2 to 5 times lower contrast between the photosphere and the starspots that are often the cause of the variability (Eiroa et al. 2002). The limits to infrared RV studies are at present unknown, but there are a number of activities underway to push down to 50 m s$^{-1}$ (veiling and fast rotation permitting!). While valuable results can be expected from near-term infrared radial velocity studies, the RV technique will likely be confined to the detection of "Hot Jupiters" with orbits inside 0.1 AU.

### D. A Survey For Young Planets

On the one hand, the search for young planets favors stars closest to the Sun, since source brightness and angular scale both improve with proximity. On the other hand, the nearest stars with ages less than 10 million years are in clusters as far away as 60-140 pc (Figure 2). Thus, samples must include classical and weak-lined T Tauri stars (ages < 10 Myr but at distances of 60-140 pc), stars in the nearby, young moving groups at 10-50 pc; Zuckerman & Song 2004) with ages from 10 Myr (when observations shows gas disks dissipate) up to 100 Myr (by which time the terrestrial planets have formed and young planetary systems should become indistinguishable from those of mature stars), and stars less than 1 Gyr when planets smaller than a few Jupiter masses contract and fade from view.

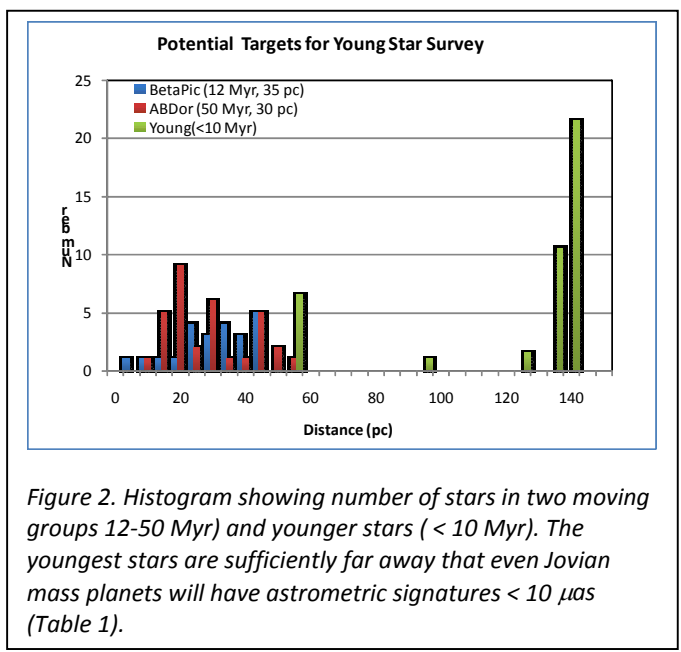

Figure 2. Histogram showing number of stars in two moving groups 12-50 Myr) and younger stars ( < 10 Myr). The youngest stars are sufficiently far away that even Jovian mass planets will have astrometric signatures < 10 µas (Table 1).

An astrometric survey of approximately 200 T Tauri stars matched to the detection of Saturn mass (or greater) planets from 1-5 AU can be accomplished in the time allocated for a SIM Lite Key Project (Table 1; Beichman et al. 2001, 2009; Tanner et al. 2007). Observations of another few hundred young stars could be added relatively inexpensively to provide astrometric data for planets detected via direct imaging or to make surveys of additional stellar types.

### III. Recommendation

Other white papers will address the progress possible with imaging and infrared radial velocity programs. In this paper we argue that an astrometric mission with the characteristics of SIM-Lite and capable of few µ-arcsec astrometry (single measurement accuracy) of V<12 mag objects over a time period of 5-10 years is necessary to provide: a) a survey of young stars, particularly the youngest stars out to 140 pc, to look for gaseous and icy planets in the critical 1-5 AU orbital range; and b) dynamical masses for gas giant planets found in imaging surveys with ground-based telescopes or with JWST. An astrometric survey of some 200 young stars could find and determine masses and orbits at least 20 (and perhaps many more) planets around stars with ages ranging from 2 to 100 Myr in orbits that span the region where planets are thought to form. An astrometric mission could also provide masses for directly imaged gas giants. The combination of astrometric and imaging data would revolutionize our understanding of planet formation in the same way that RV and transit observations are presently advancing our understanding of mature planets.

## IV. References

Baraffe, I., Chabrier, G., Allard, F., and Hauschildt, P. H., 2002 A&A, 382, 563.
Beichman, C., 2001, ASP Conference 244, (eds. Jayawardhana, R. and Greene, T.), p. 376.
Beichman et al. 2009, *The SIM Lite Astrometric Observatory*, Jet Propulsion Laboratory, http://planetquest.jpl.nasa.gov/SIM/files/Chapter-1-LR.pdf
Boss, A. P., 2001, ApJ., 562, 842.
Burrows, A. et al. , 1997, ApJ., 491, 856.
Carpenter, J. M. et al. , AJ, 121, 3160.
Chabrier et al. 2003, A&A, 402, 701.
Charbonneau, D. et al. 2007, ApJ, 658, 1322.
Chauvin, G., et al. 2005, A&A, 438, L25.
Eiroa, C. et al. , 2002, A&A 384, 1038–1049.
Fischer, D and Valenti, J., 2005 ApJ, 622, 1102.
Fortney, J. J., Marley, M. S., Saumon, D., Lodders, K. 2008, ApJ 683,1104
Huélamo, N. et al. , 2008, A&A, 489, 9.
Ida, S. and Lin, D. C., 2005, ApJ., 626, 1045.
Kalas, P. et al. , 2009, Science, 322, 1345.
Krist, J. 2007, proc. Lyot Conference, in press.
Liu, X. Burrows, A. and Ibgui, L. 2008, ApJ, 687, 1191.
Marios, C. et al. , 2009, Science, 322, 1348.
Marley, M. S., Fortney, J. J., Hubickyj, O., Bodenheimer, P., Lissauer, J. J. 2007 ApJ, 655, 541.
Millan-Gabet,. R. and Monnier, J.D. 2009, Decadal White Paper.
Neuhäuser, R., et al. 2005, A&A, 435, L13.
Oppenheimer, B and Hinkley, S. 2009, ARAA, In press.
Pollack, J. B., Hubickyj, O., Bodenheimer, P., Lissauer, J. J., & Podolak, M., 1996, Icarus, 124, 62.
Prato, L., et al. 2008, ApJ, 687, L103.
Robinson, S. 2008, PhD Thesis, UCSC.
Seager, S., Kuchner, M., Hier-Majumder, C. A., Militzer, B. 2007, ApJ, 669, 1279.
Setiawan, J., et al. 2008, Nature, 451, 38.
Sivaramakrishnan, A. et al. 2009, ApJ, 688, 701.
Skumanich, A., 1972, ApJ, 171, 565.
Tanner, A. et al. , 2007, PASP, 119, 747.
Zuckerman, B. and Song, I., 2004, ARAA, 42, 685.

## V. Acknowledgements

The research described in this paper was carried out at the Jet Propulsion Laboratory (JPL) , California Institute of Technology, under contract with the National Aeronautics and Space Administration. Work with NIRCam at JPL is supported under JPL Contract Task Plan No. 70-7920.